# A CONTROL SYSTEM OF THE JOINT-PROJECT ACCELERATOR COMPLEX

J. Chiba, H. Fujii, K. Furukawa, N. Kamikubota, H. Nakagawa, N. Yamamoto,
High Energy Accelerator Research Organization (KEK), Tsukuba, Japan 305-0801
H. Sakaki, H. Yoshikawa,
Japan Atomic Energy Research Institute (JAERI), Tokai, Japan 319-1195

## Abstract

The current status of the control system for a new high intensity proton accelerator, the JAERI-KEK Joint Project, is presented. The Phase 1 of the Joint-Project has been approved and recently started its construction at JAERI site at Tokai. The first beam commissioning is scheduled in 2006. In parallel with it, a 60-MeV Linac is now being constructed at KEK site at Tsukuba for R&D purpose.

Recently the Project has officially decided to use the Experimental Physics and Industrial Control System (EPICS). Under the EPICS environment, we are challenging to implement the Ethernet/IP network for all communication, even at the level of end-point controllers which are so far connected via a field bus. In order to realize such a system, three new controllers (PLCs, WE7000 stations and general-purpose Ethernet boards) are being developed. A prototype EPICS driver for the PLCs works fine and is used to control the ion-source at the KEK Linac.

## 1 THE JOINT-PROJECT

As reported in the previous conference [1] the two projects, JHF at KEK and NSP at JAERI, were merged and formed a new project, the JAERI-KEK Joint Project. The Phase 1 of the project consists of three accelerator facilities (400-MeV Linac, 3-GeV Rapid-cycling Synchrotron, and 50-GeV Synchrotron) and two experimental facilities (Material and Life Science Facility at 3 GeV and Nuclear and Particle Physics Facility at 50 GeV). The construction of the Phase 1 will be completed in 2006. The schedule of the Phase2 is not known yet, in which 600-MeV Super Conducting Linac and Accelerator Driven Transmutation (ADS) Experimental Facility will be constructed.

We are planning to make a central control system with which not only the accelerator but also the beams and targets in the experimental facilities can be monitored and operated in order to reduce the operational costs including labor.

In addition to the facilities at Tokai mentioned above, we have constructing a 60-Mev Linac at KEK site. It comprises an ion source, a low-energy beam transport (LEBT), 3-MeV RFQ Linac, a medium-energy beam transport (MEBT), a 50-MeV drift-tube Linac (DTL), and a 60-MeV separated DTL (SDTL). The DTL and SDTL will be transported and re-installed at Tokai after successful beam tests.

## 2 DESIGN OF THE CONTROL SYSTEM

### 2.1 Introduction

Before forming the Joint Project, the use of the EPICS toolkit at the JHF accelerator had been decided after intensive discussion among the JHF control group members [1]. Although there were some complaints against the use of EPICS, we recently decided to use it in the Joint Project. Since the Joint JAERI-KEK construction team have been formed only a half year ago, concrete rules needed for the control system design are not determined yet except for the use of EPICS.

In the following subsections, we describe our present thought on the control system in 5 components; (1) device control, (2) network, (3) operator interface, (4) interface to beam simulator and (5) database.

### 2.2 Device control

In EPICS, all device controls are carried out through input/output controllers (IOCs). At present, only VMEs under the vxWorks operating system can be used as an EPICS IOC. There are many VME modules are commercially available for processing normal digital and analogue signals. EPICS drivers are already available for many of those modules. Therefore, we don't need to worry about devices which require only those signals. However, there are many devices which need special treatments.

PLCs are cheap and reliable in a sense that they keep running under any trouble in computers or networking, and therefore it is preferable to use PLCs for some critical devices such as an ion-source. How PLCs make communication with IOCs is one of the issues we have to determine early. Our answers are as follows;

- Use Ethernet TCP/IP. Among PLCs commercially available, presently only FA-M3 PLC by Yokogawa support data transfers protocols we required, we force equipment groups to use it if they need PLCs.
- Assign an identification number (ID) to each PLC. It is dangerous to rely only on the IP address. Before any communication with a PLC and at any data (command) transfer to a PLC, check the ID to make sure you are communicating with an appropriate PLC.
- No individual commands for operation. Only data transfer from/to PLC memory is supported. Any device operation should be done though contents of memory in a restricted address range.
- No direct write to PLC memory. An IOC sends 3-word data (an ID, a memory address and content) to a communication area in the PLC memory. Then, a PLC ladder program moves the content to an appropriate address after checking its ID.

A prototype of the PLC EPICS driver has been written and successfully used for the ion-source operation at the KEK 60-Mev Linac, although all the rules mentioned above are not fully implemented yet in the prototype driver. Detailed description on the device driver may be found elsewhere [2].

Choice of waveform digitizers is another important issue we have to consider for device control. As reported in the previous conference [3], our choice was the WE7000 measurement station by Yokogawa. Its EPICS driver has been partially completed and tested. For this station, TCP/IP Ethernet communication method is used as same as that for the PLCs. Detailed description is given again in Ref. [2].

For precise current control of power supplies, we should avoid extending an analogue signal in a long distance, and instead, should use digital data links. Since Ethernet network runs all over the area, we chose Ethernet for digital data link avoiding another type of field buses so that all of the data communication is unified. By doing so, we can trace back all data through network which will significantly help us for operation diagnostics. Also it reduces the number of stock items which may reduce the operational cost. To achieve the unified Ethernet links, we are developing an Ethernet interface board which will be used primarily in the power supplies for quadrupole magnets at the DTL. The board is designed considering more general uses and may be embedded in other devices.

However, some devices required for some equipment may communicate with our control system only via a method other than Ethernet. For example, most of low-price power supplies have a GPIB or RS232c interface for remote access. In such cases, we will use Ethernet interface such as a GPIB LAN gateway for GPIB and a terminal server for RS232c to minimize the area covered by field buses.

*2.3 Network*

Redundant gigabit Ethernet with fiber optics linked like as a "star" starting from the central control room (CCR) to each accelerator (Linac, 3-GeV RCS and 50-GeV PS) and experimental facility is the backbone of the control network. From each backbone station, star-like 100baseFX optical fiber cables extend to network switches over the facility. Most of the network end nodes such as IOCs are 100baseTX. Some nodes in a high EMI noise levels such as near klystrons or high-powered pulsed power supplies should be 100baseFX.

As described in previous subsection, we are going to use Ethernet links in stead of other field buses. Therefore the total number of IP addresses required for the control system could easily exceed a few thousands. Therefore it is impossible to use a global IP address space. Instead we are going to use a Class B private IP address space and use a router to communicate outside the network.

We have not figured out how the network should be divided into subnets. The EPICS channel access (CA) protocol requires the network broadcast. Therefore all EPICS nodes should be in one subnet. Nodes for diagnostic purpose should belong to a different subnet because they must be free from the EPICS subnet. Such diagnostic nodes include a communication port of a traffic watcher (of course, it should have also a port in the EPICS subnet to spy network traffic) and an IOC's CPU console (RS232c through a terminal server). Those two types of nodes are in general close each other in space, and therefore it is better to implement the virtual-LAN technique to reduce cabling and number of network switches.

*2.4 Operator interface*

We will use one of the standard EPICS OPI tools for the EPICS operation, but not specified yet which one should be used. Also there are two choices; (1) OPI applications run directly on a PC Linux or (2) OPI applications run on an X-client machine with an X-terminal (or a PC with X-server). The former choice is good for network traffic but not good for maintaining applications. The latter is just opposite. Another type of operator interface is necessary which is related to the issues described in the next subsection.

### 2.5 Interface to accelerator simulator

In order to operate the accelerator stably, it is important to simulate the beam before parameters are physically changed. Accelerator physicists are responsible for developing such a simulation code, but control group should help them to include the EPICS interface and operator interface in the code. SAD code [4] used at the KEKB accelerator is a candidate in the Joint-Project. Since SAD code used at KEKB already has the graphical user interface (GUI) and the EPICS interface, we don't need to worry so much about the interfaces if SAD is chosen as a simulation code. However, SAD must be significantly modified or new simulation code should be implemented for the Linac simulation. If new code is chosen, we have to provide the GUI and EPICS interface similarly to the SAD case.

### 2.6 Database

For steady and reliable operations of the accelerator, use of proper database system is indispensable. We have just started to design various kinds of database necessary in our system.

## 3 PRESENT STATUS

At KEK, an EPICS application server (HP-UX, HP 9000 model D380/2) and several VME IOCs (Force PowerCore 6750) have been installed primarily for software developments.

Development of an EPICS device driver for the WE7000 measurement station (Yokogawa) is being carried out. We have so far developed EPICS device support for 3 WE7000 modules; WE7111 (100MS/s digital oscilloscope), WE7121 (10MHz function generator) and WE7271 (4ch 100kS/s waveform digitizer). It works basically fine. However, concurrent use with MS Windows applications doesn't work with unknown reasons. Because the WE7000 have been developed for MS Windows OS, there are fruitful applications available under Windows system. Those applications are very useful for maintenance of the WE7000 stations and developments of EPICS applications. We must solve this problem soon because the concurrent use with Windows is indispensable.

Present version of the EPICS driver for Yokogawa's FA-M3 PLC supports only simple input/output operations. The rules on the PLC use mentioned in Sec. 2 are not implemented yet.

The EPICS system will be used also for the operation of the 60-MeV Linac. At present, only the ion-source control which uses a MA-F3 PLC is under the EPICS environment.

The HP EPICS server, VME IOCs, PLCs and 3 operators PC terminals form a Class C network with private IP address. Operator can access to KEK-LAN via routers (2 of freeBSD PCs). The HP server also has a network port to KEK-LAN through which we access to the system from outside of the accelerator building.

At JAERI, an HP EPICS application server and several VME IOCs are installed. A test stand for klystron power supplies employs the EPICS system. An EPICS driver for a VME pulse motor controller module is now under development.

## 4 ACKNOWLEDGEMENTS

The authors would like to thank all members of the Joint-Project team. They also acknowledge members of the KEKB ring control group for their advices and suggestions.